 \documentclass[epj]{svjour}
\usepackage{amsmath, amssymb, amstext}
\usepackage{bbm}
\usepackage{booktabs}
\usepackage{array}
\usepackage{overpic}
\usepackage{pdflscape}
\usepackage{longtable}
\usepackage{caption}

\usepackage[varg]{txfonts} 
\usepackage[latin1]{inputenc}

\newcommand{\Om}{{\Omega}}
\newcommand{\anti}[1]{\overline{#1}}
\begin{document}
\title{%
Phenomenological view on baryon-baryon potentials from lattice QCD simulations 
}%
\author{
J.~Haidenbauer\inst{1}  \and Ulf-G. Mei\ss{}ner \inst{2,1,3} 
}


\institute{
Institute for Advanced Simulation and J\"ulich Center for Hadron Physics,
Institut f\"ur Kernphysik, Forschungszentrum J\"ulich, D-52425 J\"ulich, Germany
\and
Helmholtz-Institut f\"ur Strahlen- und Kernphysik and Bethe Center
for Theoretical Physics, Universit\"at Bonn, D-53115 Bonn, Germany
\and
Tbilisi State University, 0186 Tbilisi, Georgia
}
\abstract{ 
A qualitative discussion on the range of the potentials as they result from the phenomenological 
meson-exchange picture and from lattice simulations by the HAL QCD Collaboration is presented. For 
the former pion- and/or $\eta$-meson exchange are considered together with  the scalar-isoscalar 
component of correlated  $\pi\pi /K \bar K$ exchange. 
It is observed that the intuitive expectation for the behavior of the baryon-baryon potentials for 
large separations, associated with the exchange of one and/or two pions, does not always match with 
the potentials extracted from the lattice simulations.
Only in cases where pion exchange provides the longest ranged contribution, like in
the $\Xi N$ system, a reasonable qualitative agreement between the phenomenological and the lattice 
QCD potentials is found for baryon-baryon separations of
$r \gtrsim 1$~fm. For the $\Omega N$ and $\Omega\Omega$
interactions where isospin conservation rules out one-pion exchange a large mismatch
is observed, with the potentials by the HAL QCD Collaboration being much longer ranged and 
much stronger at large distances as compared to the phenomenological expectation.
This casts some doubts on the applicability of using these potentials in few- or many-body systems. 
\PACS{
        {12.38.Gc}{} \and 
        {13.75.Cs}{} \and 
        {13.75.Ev}{} \and 
        {14.20.-c}{} 
      }
} 
\maketitle
\section{Introduction} 
\label{sec:Intro} 

The study of the strong interaction as given by the 
fundamental theory of QCD on the lattice has made significant
progress over the last few years, not least due to the
availability of high performance computers and improved algorithms. This progress
is documented in several review articles, see, e.g., 
Ref.~\cite{Fodor:2012} as far as the masses of light hadrons 
are concerned, or Refs.~\cite{Briceno:2017} and \cite{Aoki:2012,Beane:2010,Davoudi:2018}
with regard to the scattering of two mesons or two 
baryons, respectively. 

However, in recent times there has been also some discord. 
This concerns in particular baryon-baryon scattering, where there is an ongoing 
controversy \cite{Iritani:2016,Iritani:2017,Beane:2017,Davoudi:2017,Iritani:2018A}  
about the applicability of one of the basic tools of lattice QCD (LQCD),  
namely the L\"uscher finite volume formula \cite{Luscher:1986,Luscher:1991}
which is commonly used to relate the energy 
levels obtained in LQCD simulations to two-body phase shifts. 
In addition, there have been criticial remarks on the so-called
HAL QCD method \cite{HALQCD:2012}, suggested in Ref.~\cite{Iritani:2017}
as a way to circumvent the difficulties with the L\"uscher approach. 
In this method the Bethe-Salpeter wave function is extracted
from the lattice simulation and based on it, a (local) potential
is constructed which is then utilized to calculate the phase shifts.
This method was called into question very recently, see the discussions in 
Refs.~\cite{Yamazaki:2017,Aoki:2017,Yamazaki:2018}.
Note that there are actually two different methods proposed 
and employed by the HAL QCD Collaboration, a time-dependent one 
and the more recent time-independent one \cite{HALQCD:2012}, 
also called imaginary-time HAL QCD method. 

In the present work we do not dwell into formal aspects and, specifically,
we do not add anything directly to those controversies mentioned above.  
Rather we would like to concentrate on the intuitive
and phenomenological side, namely on the potentials themselves
as they emerge from the calculations and publications of the
HAL QCD Collaboration. Of course, potentials are not observable 
physical quantities, see e.g. the discussion in the review~\cite{Epelbaum:2008ga}.
Nonetheless, since the days of  Yukawa~\cite{Yukawa:1935xg}, 
potentials (in configuration space) have played an important role 
as an intuitive visual guidance for the interpretation of the reaction 
dynamics. For example, the nucleon-nucleon ($NN$) interaction, as the 
most prominent case, is seen as being composed of a long-range part 
provided by pion exchange, an attractive intermediate-range part that is 
due to (correlated) $\pi\pi$ exchange \cite{Lacombe:1980,Reuber:1996}
often represented by a scalar meson called $\sigma$ or $f_0$(500) \cite{PDG}, 
and, finally, a repulsive short-range part that is due to vector-meson 
exchange, specifically the $\omega$ \cite{Machleidt:1987}. 

Certainly, this traditional view might have been one of the reasons why 
the work of the HAL QCD Collaboration focusses prominently on potentials. 
But do the potentials extracted from the lattice simulations indeed meet 
the intuitive expectations formulated above? In particular, do they exhibit 
the features we would anticipate from the meson-exchange dynamics? 
These are the questions we want to address in the present
study. Thereby, we concentrate on the long range behavior of 
the potentials, i.e. on the results for baryon-baryon distances $r\gtrsim 1$~fm. 
Obviously, only for large separations the dynamics can be expected to be
simple enough and accessible for an intuitive interpretation. 

With that aim in mind we take a closer look at the potentials in the 
$NN$, $\Xi N$, $\Om N$, and $\Om\Om$ channels. For several $S$-wave states 
of those systems lattice simulations by the HAL QCD Collaboration are
available, performed at almost physical masses ($m_\pi = 146$ MeV). 
Pertinent results for the potentials can be found
Ref.~\cite{Sasaki:2018} for $\Xi N$ ($^3S_1$; $I=0$),
in \cite{Iritani:2018} for $\Om N$ ($^5S_2$),
and in \cite{Gongyo:2017} for $\Om\Om$ ($^1S_0$).
Like in case of $NN$, pion exchange provides the long-range contribution
to the $\Xi N$ potential, supplemented by correlated $\pi\pi$ ($\sigma$)
exchange at somewhat shorter distances. The situation is different for $\Om N$ and $\Om\Om$. 
Since the isospin of the $\Om$ baryon is $I=0$, pion exchange is not allowed
by conservation of isospin and should be strongly suppressed. The 
contribution with the longest range should be due to $\eta$ exchange. 
Also correlated $\pi\pi$ exchange should be suppressed because, again for
isospin reasons, the $\pi\pi$ state cannot couple directly to the $\Om$ 
but only via $K\bar K$ and/or $\eta\eta$ \cite{Reuber:1996,Sekihara:2018}.
This is different from the situation in $\Lambda N$ or $\Lambda \Lambda$, say, 
where leading-order pion exchange is likewise not possible. 
However, two-pion exchange can contribute due to the coupling to $\Sigma N$ or
$\Sigma \Sigma$ \cite{Reuber:1996}. In case of $\Om$ there is no other baryon
with the same strangeness quantum number that would facilitate such a coupling.

In the present paper we evaluate the potentials for the baryon-baryon
channels in question based on pion- or $\eta$ exchange and supplement them
with contributions from correlated $\pi\pi / K\bar K$ exchange in the
scalar-isoscalar (i.e. ''$\sigma$'') channel with coupling strengths taken
from microscopic models \cite{Reuber:1996,Sekihara:2018}. The resulting
potentials are then confronted with the ones that are extracted by the HAL
QCD Collaboration from their lattice simulations. As we will see, there is
a reasonable qualitative agreement in case of the $\Xi N$ interaction. 
However, for the $\Om N$ and $\Om\Om$ interaction we observe a striking
difference. Here the potentials from lattice QCD are much longer ranged
and much stronger for large baryon-baryon separations than what one would
expect from the meson-exchange picture. 

The paper is structured as following:
The basic ingredients of our calculation are summarized in Sect.~2. 
A brief overview of the evaluation of the potential from correlated $\pi\pi / K\bar K$ 
exchange via dispersion theory is provided in an appendix. 
Our results are presented in Sect.~3. We first discuss the interactions
in the $NN$ and $\Xi N$ channels where the long-range part is provided
by pion exchange. Then we consider the $\Om N$ and $\Om\Om$ interactions 
where only $\eta$ exchange or even shorter-ranged contributions are possible. 
The paper closes with concluding remarks. Some technicalities are relegated to an Appendix.

\section{Ingredients} 
\label{Sec2} 

The potentials resulting from the exchange of a pseudoscalar (ps) meson and a scalar (s) 
meson between the baryons $B$ and $B'$ for the $^1S_0$ ($^5S_2$) partial wave are given 
by \cite{Machleidt:1987}
\begin{eqnarray}
\label{Vpi}
V_{ps} &=& \frac{1}{3}\frac{f_{BB{ps}}f_{B'B'{ps}}}{4\pi} m_{ps} \, Y(m_{ps} r) \ {\boldmath\mathcal{O}}, \\
V_{s}  &=& -\frac{g^2_{s}}{4\pi} m_{s} \left[1 - \frac{m_{s}^2}{4 M_B M_{B'}}\right] \, Y(m_{s} r),
\label{Vsi}
\end{eqnarray}
where $Y(x) = e^{-x}/x$. $m_{ps}$ and $m_s$ stand for the masses of the mesons. 
The potential for ps exchange needs to be multiplied with the expectation values for
the appropriate spin 
(\mbox{\boldmath $\sigma_1\cdot\sigma_2$}, 
\mbox{\boldmath $\Sigma_1\cdot\sigma_2$}, 
\mbox{\boldmath $\Sigma_1\cdot\Sigma_2$}) 
and/or isospin operators, see Ref.~\cite{Haidenbauer:2017}, indicated in Eq.~(\ref{Vpi}) symbolically
by \mbox{\boldmath $\mathcal{O}$}.  
There is an additional term involving the (irreducible) tensor operator in case of ps exchange, 
and one involving the spin-orbit operator in case of scalar exchange \cite{Machleidt:1987}.  
But their expectation values vanish for the $S$-wave states considered and, therefore, they 
are omitted in Eqs.~(\ref{Vpi}) and (\ref{Vsi}). 

With regard to the coupling constants of the ps mesons in Eq.~(\ref{Vpi}) we note that
$f_{NN\pi} / m_\pi  = g_A /(2 f_\pi)$, where $g_A$ is the axial-vector strength and $f_\pi$
the weak pion decay constant. In the actual calculation the values 
$f_\pi = 92.1$~MeV and $g_A = 1.26$ \cite{PDG} are used. 
The strength for the other couplings of ps mesons to octet baryons are fixed by imposing
SU(3) flavor symmetry \cite{Haidenbauer:2013} based on $f \equiv g_A / (2 f_\pi)$ 
(see Table~\ref{tab:1}) substituting, however, the physical $\eta$ decay constant, 
$f_\eta =1.3 f_\pi$, in the actual calculations.
 
For the coupling of ps mesons to decuplet baryons we proceed in the same way, see 
Ref.~\cite{Haidenbauer:2017}. There are estimates for the corresponding coupling constants 
$g_1$, relevant for $f_{\Delta\Delta\pi}$, in the non-relativistic quark model \cite{Brown}
and from large $N_c$ considerations \cite{Fettes:2000,Zhu}
which lead to $g_1 \approx 9/5\, g_A$, i.e. $g_1 \approx 2.27$.
On the other hand, lattice QCD calculations suggest $g_1 \approx g_A/ 2$ ($g_1 \approx 0.60$) 
\cite{Alexandrou:2013}. Note that the decuplet-decuplet coupling constant $f_{DD}$ used in
Table~\ref{tab:1} is $f_{DD} \equiv g_1 / (2 f_\pi) / 9 $, due to our normalization of the 
isospin-3/2 operators \cite{Haidenbauer:2017,Wiringa:1984}, so that 
$f_{\Delta\Delta\pi}=f_{NN\pi}/5$ for the quark-model value.  

\begin{figure}
\centering
\includegraphics[width=7.0cm,angle=-90]{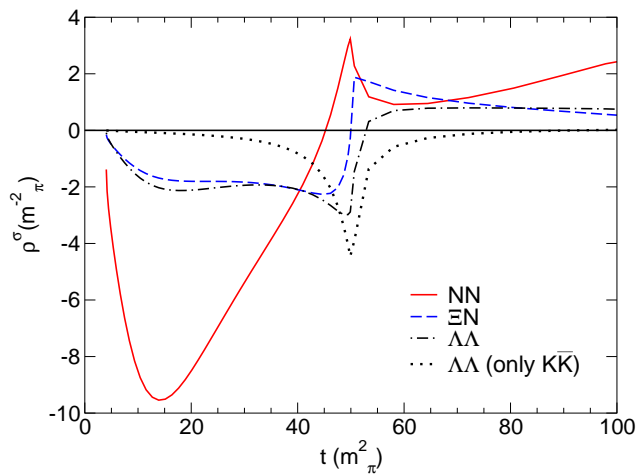}
\caption{Spectral function $\rho^\sigma (t)$ for the scalar component of correlated
$\pi\pi$-$K\bar K$ exchange in the scalar-isoscalar channel of various baryon-baryon
channels (cf. Fig.~16 in Ref.~\cite{Reuber:1996}): 
$NN$ (solid line), $\Xi N$ (dashed line), $\Lambda\Lambda$ (dash-dotted line).
The dotted line indicates the results for $\Lambda\Lambda$ when only $K\bar K$
exchange is kept in the Born term. 
} 
\label{fig:spec}
\end{figure}

The coupling constants for the scalar ($\sigma$ or $f_0$(980)) exchange are estimated 
from a microscopic calculation of correlated $\pi\pi/K\bar K$ exchange. 
A corresponding model study for baryon-baryon ($BB$) systems has been presented in 
Ref.~\cite{Reuber:1996} for the ``$\sigma$'' (i.e. for the $J=0$, $I=0$ channel) as well 
as for the ``$\rho$'' (i.e. the $J=1$, $I=1$  channel). 
It is based on dispersion theory and utilizes crossing symmetry
to relate the correlated $\pi\pi/K\bar K$ exchange with
amplitudes in the crossed channels $B\bar B \to \pi\pi,K\bar K$.
For a comprehensive description of this model we refer the reader to 
Ref.~\cite{Reuber:1996}. A short overview of the basic ingredients and details on 
how the coupling constants are obtained are provided in the appendix. 
The effective $\sigma$ coupling strengths we employ are taken from Table~5 of 
Ref.~\cite{Reuber:1996}. 
The underlying spectral functions $\rho^{\sigma}$ from which those coupling constants are 
extracted are reproduced in Fig.~\ref{fig:spec}.
The actual value of the coupling constants are $g_s^2/4\pi = 7.77$ for $NN$ scattering 
and $g_s^2/4\pi = 1.52$ for $\Xi N$. They are based on an effective $\sigma$ mass of 
$550$ MeV \cite{Reuber:1996}, a choice suggested by the behavior of $|\rho^{\sigma}(t)|$ 
for $NN$ which is strongly peaked around this value, cf. Fig.~\ref{fig:spec}. 

The effective $\sigma$ coupling strength for the $\Om\Om$ channel has not been
evaluated in Ref.~\cite{Reuber:1996}. Is strength depends crucially on the coupling 
constants of the ps mesons to the decuplet baryons which are quite uncertain,
as mentioned above. In view of that we refrain from attempting an explicit 
but tedious evaluation of the contribution for correlated $2\pi$ exchange. 
Rather we aim for a qualitative estimation of the coupling strength and we
focus primarily on the expected range, where pertinent information can be 
deduced by considering the situation in the $\Lambda\Lambda$ system. 
For the latter the corresponding spectral function has been calculated in
\cite{Reuber:1996} and it is shown in Fig.~\ref{fig:spec} (dash-dotted line). 
The deduced effective $\sigma$ coupling strength is $g_s^2/4\pi = 2.00$.
Indeed, since the small $t$ behavior for $\Xi N$ and $\Lambda\Lambda$ are similar 
one expects a $\sigma$-like interaction with comparable strength and range 
in both channels.
  
\begin{figure}
\vglue 7.3cm
\includegraphics{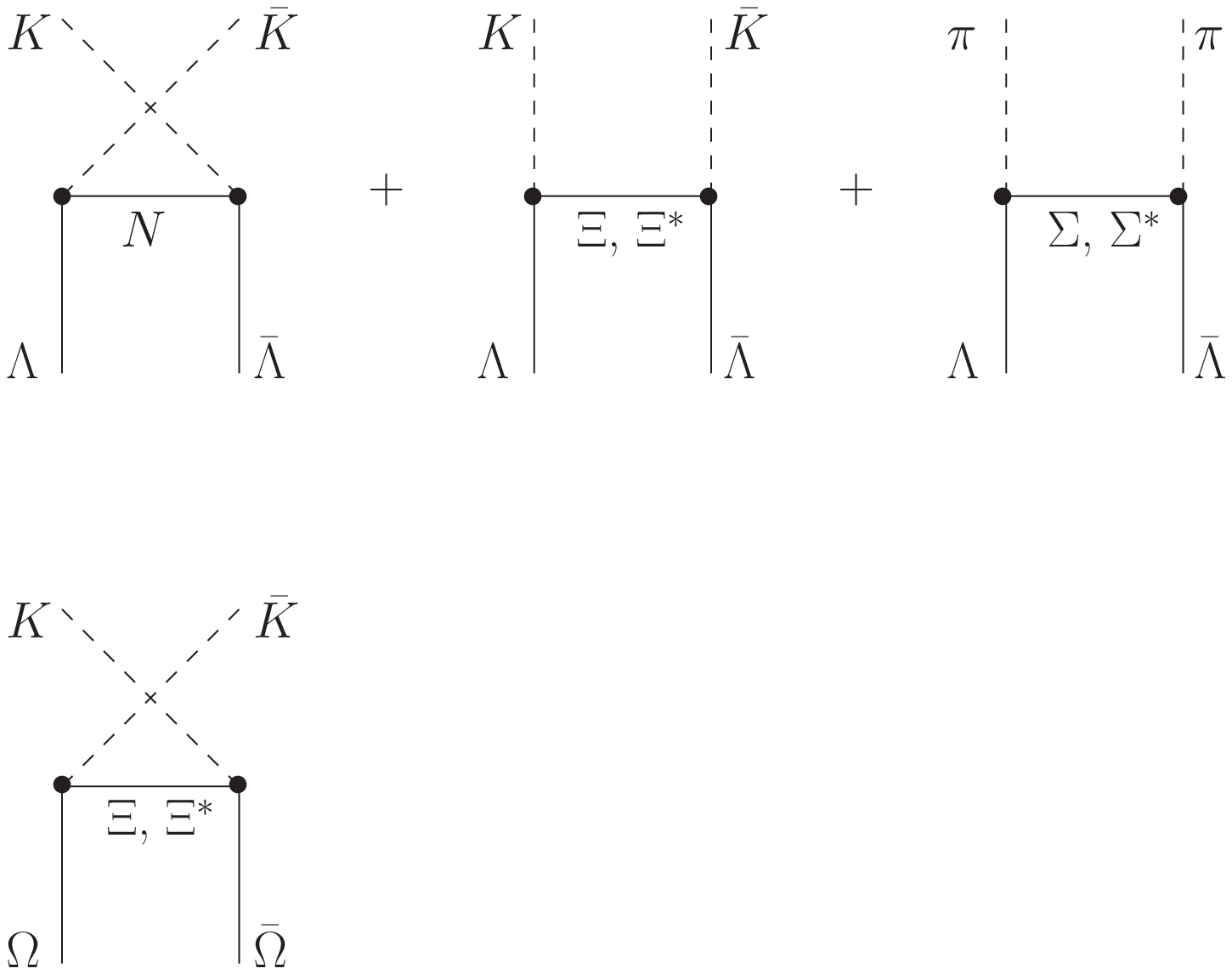}
\caption{The contributions to the Born amplitude ($V_{B\bar B' \to \alpha}$ 
in Fig.~\ref{Tran}) for correlated $\pi\pi/K\bar K$ exchange in case of the
$\Lambda \Lambda$ (top panel) and $\Om\Om$ (bottom panel) interactions.
}
\label{Born}
\end{figure}

When taking those results as guideline for $\Om\Om$ one has to keep in mind, 
however, that the spectral function for $\Lambda\Lambda$ 
receives contributions from several pieces as depicted in Fig.~\ref{Born}. 
Some of those involve a coupling to $\pi\pi$ (in combination with $\Sigma$, $\Sigma^*$ 
states), and those are the ones which give rise to the long range part of correlated 
$\pi\pi/K\bar K$ exchange. Others involve only a coupling to $K\bar K$ (combined with
a nucleon or $\Xi$, $\Xi^*$) and they provide only shorter ranged 
contributions. In case of correlated $\pi\pi/K\bar K$ exchange in $\Om\Om$ 
scattering only kaons in combination with $\Xi$, $\Xi^*$ states can contribute, 
see Fig.~\ref{Born}. 
We can easily simulate this situation by re-calculating the spectral function for 
$\Lambda\Lambda$ with the $\Sigma$, $\Sigma^*$ exchanges switched off. 
The corresponding $\rho^\sigma$ is indicated by the dotted line in Fig.~\ref{fig:spec}. 
Clearly, now the small $t$ (long range) part is depleted and the spectral function 
is basically concentrated around the $K\bar K$ threshold. Transformed to 
$r$-space it is best represented by the exchange of the $f_0$(980) meson 
whose effective coupling constant can be deduced from the spectral function  
and amounts to $g_s^2/4\pi = 1.77$.

We will use the above strength for the effective $f_0(980)$ exchange as basis 
for the $\Omega\Omega$ interaction. Actually, one can do a simplistic 
estimation of what do expect for $\Omega\Omega$ based on the finding that
the $f_0(980)$ for $\Lambda\Lambda$ results primarily from the contribution 
from $N$ exchange. 
Considering the SU(3) structure and taking the quark-model values for the 
coupling constants $f_{N\Delta\pi}$ and $f_{\Delta\Delta\pi}$ one arrives at 
$f^2_{\Om\Xi K} = 9/2\, f^2_{\Lambda N K}$ and 
$f^2_{\Om\Xi^* K} = 6/25\, f^2_{\Lambda N K}$ \cite{Haidenbauer:2017}
which suggests that $\Xi$ exchange should dominate. Since the coupling
constants enter with the fourth power into the spectral function/potential the
effective $f_0(980)$ coupling strength could be enhanced by as
much as $(9/2)^2 \approx 20$ as compared to the $\Lambda\Lambda$ case.
Of course, one should not forget that the actual results for $\Om\Om$ will 
not only depend on the relevant ($\Om\Xi K$, $\Om\Xi^* K$) coupling constants, 
but there is also a different spin-momentum structure at the vertices, a 
difference in the masses of the involved baryons as compared to the $\Lambda\Lambda$ 
case, etc. 

\begin{table}[h]
\renewcommand{\arraystretch}{1.6}
\centering
\caption{SU(3) relations for the relevant coupling constants.
For the $F/(F+D)$-ratio we adopt the value $\alpha=0.4$. 
}
\label{tab:1}
\vskip 0.2cm 
\begin{center}
\begin{tabular}{ c }
\hline
{octet baryons} \\
\hline
$f_{NN \pi}= f$ \qquad $f_{\Xi \Xi \pi} = -(1- 2\alpha) f$ \qquad
$f_{NN \eta}= \frac{1}{\sqrt{3}}(4\alpha -1) f$ \ \\
\hline
{decuplet baryons} \\
\hline
\ $f_{\Delta\Delta \pi}= f_{DD}$ \qquad $f_{\Om \Om \eta} = -\sqrt{12} f_{DD}$ \\  
\hline
\end{tabular}
\end{center}
\end{table}
\renewcommand{\arraystretch}{1.0}

\section{Results} 

\subsection{Results for $NN$ and $\Xi N$} 

\begin{figure}
\centering
\includegraphics[width=7.0cm,angle=-90]{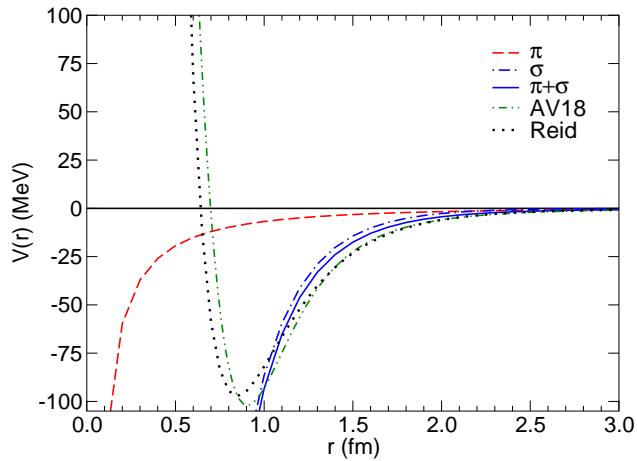}
\caption{Potential for $NN$ scattering. 
The $^1S_0$ partial wave with isospin $I=1$ is shown. 
The dashed line is the contribution from pion exchange,
the dash-dotted one is from correlated $\pi\pi-K\bar K$
($\sigma$) exchange and the solid line is their sum.
The dotted and dash-double-dotted curves represent the results for the
Reid \cite{Reid:1968} and Argonne V18 \cite{Wiringa:1995}
potentials, respectively. 
} 
\label{fig:NN}
\end{figure}

Let us start with a pedagogical case, namely the $NN$
potential in the $^1S_0$ partial wave. In this case the 
wealth of scattering data has allowed to pin down the 
phase shifts rather precisely and, as a consequence, 
the $r$-dependence of (local) potentials that reproduce 
those shifts is rather well constrained. This is illustrated 
in Fig.~\ref{fig:NN} where results for the Reid \cite{Reid:1968} 
(dotted line) and Argonne V18 \cite{Wiringa:1995} 
(dash-double-dotted line) potentials are presented. 
Both potentials include the contribution from one-pion exchange
(OPE) that provides the longest ranged piece of the interaction.
However, it is obvious from the figure that in the range of 
$r\approx 1-2$~fm where the essential intermediate
attraction comes from, the contribution from OPE is basically
negligible. That large attractive contribution is usually
attributed to (correlated) $\pi\pi$ exchange, often represented
by $\sigma$ exchange. 
The dash-dotted line in Fig.~\ref{fig:NN} is the contribution
from correlated $\pi\pi$ exchange based on the effective 
$NN\sigma$ coupling deduced in Ref.~\cite{Reuber:1996}, cf. 
Table~5. Note that the value suitable for OBE exchange is taken. 
The solid line is the sum of $\pi$ and $\sigma$ exchange.
One can see that this sum explains rather well the behavior 
of the potentials obtained from fitting to the phase shifts in the 
range $r\approx 1-2$~fm. This concerns not only the magnitude (strength) 
but also the shape (range), which is quite remarkable.  
At shorter internuclear distances additional dynamics becomes
relevant so that deviations have to be expected. Note also that 
no cutoff or regularization is applied to the $\pi$ and $\sigma$ 
exchange potentials shown. 

\begin{figure}
\includegraphics[width=7.0cm,angle=-90]{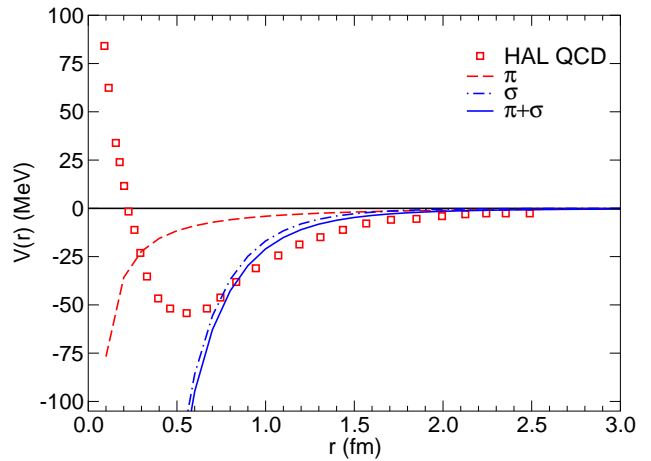}
\caption{Potential for $\Xi N$ scattering. The $^1S_0$ partial wave 
with isospin $I=0$ is shown. 
The dashed line is the contribution from pion exchange,
the dash-dotted one is from correlated $\pi\pi-K\bar K$
($\sigma$) exchange and the solid line is their sum.
Squares are results from the HAL QCD Collaboration \cite{Sasaki:2018}  
for the sink-source time-separation of $t=11$. 
} 
\label{fig:XN}
\end{figure}

In Fig.~\ref{fig:XN} corresponding results for the $\Xi N$
interaction are presented. There is no quantitative experimental 
information on $\Xi N$ scattering and, therefore, no empirical 
constraint on the potential. However, an effective $\Xi N$
has been published by the HAL QCD Collaboration \cite{Sasaki:2018}
based on a lattice simulation close to the physical point
which is included in the figure. 

Pion exchange provides again the longest ranged contribution.
The curve shown in Fig.~\ref{fig:XN} is based on a coupling 
constant fixed via SU(3) flavor symmetry which implies that
the strength is only about 20\% of that in the $NN$ channel.
There is also a contribution from correlated $2\pi$ exchange where
again the effective $\sigma$ coupling strength can be taken from 
Ref.~\cite{Reuber:1996}. 
The emerging picture is similar to that in $NN$. Specifically, at 
distances around $r\approx 1$~fm correlated $2\pi$ exchange
provides the dominant contribution while that of OPE is
small. Interestingly, the computed potential agrees fairly
well with that extracted from the lattice simulations, at
least on a qualitative level. Indeed, by re-adjusting the strength
of the effective $\sigma$ exchange by a factor of roughly two
the two potentials practically coincide in the region of 
$r\approx 1-2$~fm, suggesting that the ranges are indeed compatible. 

\subsection{Results for $\Om N$ and $\Om\Om$} 

\begin{figure}
\centering
\includegraphics[width=7.0cm,angle=-90]{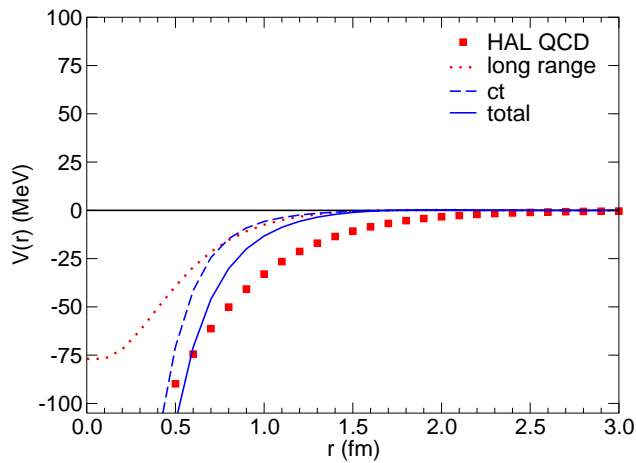}
\caption{Potential for $\Om N$ scattering. The $^5S_2$ partial wave is shown. 
Filled squares are results from the HAL QCD Collaboration \cite{Iritani:2018}
The lines are from a study within meson-exchange taken from Ref.~\cite{Sekihara:2018}. 
The dotted line is the long-ranged piece ($\eta$ plus correlated $\pi\pi-\eta\eta- K\bar K$ exchange),
the dashed line ($ct$) is a (short-range) contact term, and the solid line is the
total potential. 
} 
\label{fig:NO}
\end{figure}

For $\Om N$ scattering a full-fledged meson-exchange potential is available in
the literature \cite{Sekihara:2018}. Therefore, it is preferable to utilize 
that one directly for the comparison with the potential extracted from the lattice 
simulations, specifically because that potential is constructed in such a way that
it reproduces the HAL QCD results at low energies (scattering length, bound state) 
in the relevant $^5S_2$ partial wave. 
The potential includes $\eta$ exchange and the possible coupling of $\Om N$ to
the $\Lambda\Xi$, $\Sigma\Xi$, and $\Lambda\Xi^*$ channels \cite{Sekihara:2018}. 
Moreover, an elaborate evaluation of the contribution from correlated $2\pi$ exchange to
the $\Om N$ potential has been performed, which involves besides $\pi\pi$ and $K\bar K$
correlations also those from the $\eta\eta$ channel. These components constitute the
long-range part of the potential, indicated by the dotted line in Fig.~\ref{fig:NO}. 
In addition a contact term is included to parameterize the short-range physics,
whose range is determined by the form factor and, specifically, by the chosen cutoff 
mass of $\Lambda = 1$~GeV. Its contribution is indicated by the dashed line in 
Fig.~\ref{fig:NO} whereas the total potential is given by the solid line. 
Evidently, the so-called long-range and short-range components give rise to
rather similar contributions in the region of $r\approx 0.7-1.5$~fm. But this 
can be understood if one recalls that the ranges are set by roughly
two times the $K$ or $\eta$ mass for the meson-meson correlations and by the 
cutoff mass for the contact term, which are both in the order of $1$~GeV. 
Much more conspicuous is the difference to the potential extracted from the lattice 
simulations. The latter is clearly longer ranged and there is also more strength 
located at large distances. 

\begin{figure}
\centering
\includegraphics[width=7.0cm,angle=-90]{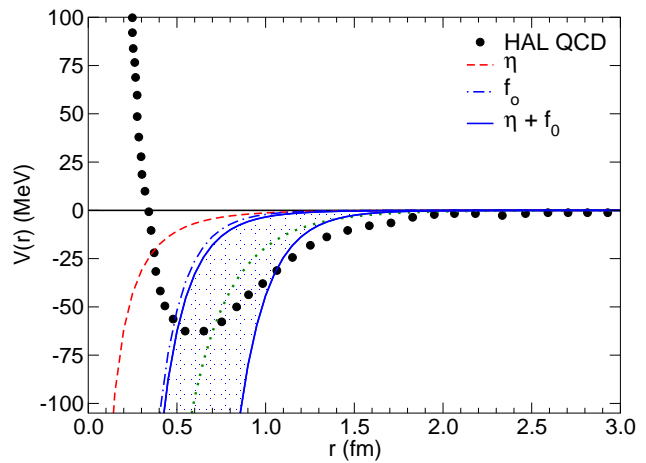}
\caption{Potential for $\Om \Om$ scattering. The $^1S_0$ partial wave is shown. 
The dashed line is the contribution from $\eta$ exchange with coupling strength
inferred from LQCD \cite{Alexandrou:2013}, the dash-dotted line is from an 
effective $f_0(980)$ exchange based on the spectral function for $\Lambda\bar\Lambda\to K\bar K$. 
The solid line is their sum with the band representing the uncertainty in the
effective $\Om\Om f_0$ coupling strength, see text.  
The dotted line is the potential from $\eta$ exchange with coupling strength 
taken from the quark model. 
Circles are results from the HAL QCD Collaboration \cite{Gongyo:2017}. 
} 
\label{fig:OOf}
\end{figure}

This peculiar feature is incorporated in the parameterization of the $\Om N$ potential 
in Ref.~\cite{Iritani:2018} by including a term with the range $2 m_\pi$, 
called (Yukawa)$^2$. As motivation for that a possible OZI violating vertex is 
quoted that should allow two pions to couple to the $\Om$. 
However, in our opinion to speak only of OZI violation in this context might be
somewhat misleading. First and foremost it is a violation of isospin symmetry 
which is required for coupling a pion (or two) to the $\Omega$. 
Indeed, the mixing of $\eta$ and $\pi^0$ provides such a contribution but, 
of course, it is expected to be rather small. For example, utilizing 
the electromagnetic mass matrix,
$$\langle \pi^0|\delta m^2|\eta\rangle = [m^2_{\pi^0}-m^2_{\pi^+}+m^2_{K^+}-m^2_{K^0}]/\sqrt{3}~,$$
as a measure for the mixing strength (see Ref.~\cite{Dalitz:1964}) 
one obtains 
\begin{equation}
f_{\Om\Om\pi} = -
\frac{\langle \pi^0|\delta m^2|\eta\rangle}{m^2_{\eta}-m^2_{\pi^0}}\, 
f_{\Om\Om\eta} \approx 0.0106\, f_{\Om\Om\eta} \ . 
\end{equation}

Results for the $\Om\Om$ potential in the $^1S_0$ partial wave are presented 
in Fig.~\ref{fig:OOf}. 
Since the coupling constants of pseudoscalar mesons to decuplet baryons is not
constrained empirically, we consider two cases: 
(a) The coupling constant from lattice simulations \cite{Alexandrou:2013} which
indicate that $f_{\Delta\Delta\pi} \approx f_{NN\pi}/20$ for the normalization of 
the spin and isospin operators used by us \cite{Haidenbauer:2017}. 
(b) The coupling strength that
results from the non-relativistic quark model (or from large $N_c$ arguments) 
which implies $f_{\Delta\Delta\pi} = f_{NN\pi}/5$.
The value for
the $\Om\Om\eta$ coupling, relevant here, is obtained from the standard 
SU(3) relations \cite{Haidenbauer:2017}, see Table~\ref{tab:1}, under
the assumption that $\eta \approx \eta_8$.
 
\begin{figure}
\centering
\includegraphics[width=7.0cm,angle=-90]{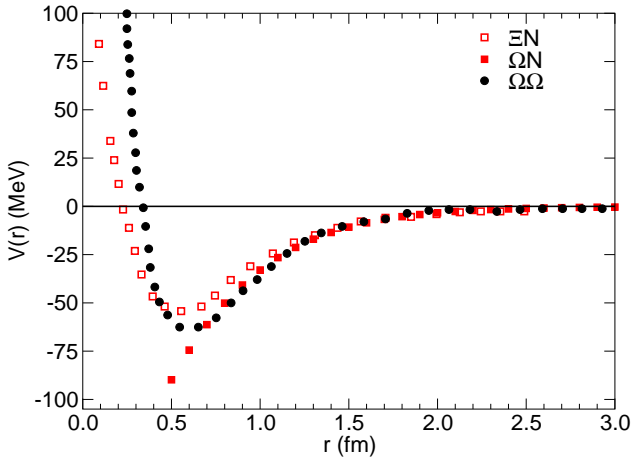}
\caption{Comparison of potentials from the HAL QCD Collaboration for
$\Xi N$ ($^1S_0$, $I=0$), opaque squares \cite{Sasaki:2018},
$\Om N$ ($^5S_2$), filled squares \cite{Iritani:2018}, and 
$\Om\Om$ ($^1S_0$) circles \cite{Gongyo:2017}. 
} 
\label{fig:HAL}
\end{figure}

The $\Om\Om$ potential that results from $\eta$ exchange with the coupling 
constant from LQCD is presented in Fig.~\ref{fig:OOf} by the dashed line.
The potential from the effective $f_0$(980) exchange (dash-dotted line) is the one 
with the strength deduced from the correlated $K\bar K$ exchange in the $\Lambda\Lambda$ 
system, cf. Sect.~\ref{Sec2}. 
Adding the two contributions and assuming that the actual $f_0$ coupling in the $\Om\Om$ 
case could be larger by a factor of up to $(9/2)^2$, as discussed in Sect.~\ref{Sec2}, 
leads to the band confined by the two solid lines. 
Thus, considering the sizable uncertainty in the effective $f_0$ coupling constant, 
the strength of the potential from LQCD at $r\approx 1$~fm can be roughly reproduced. 
However, it is obvious that the $r$-dependence deduced from the lattice simulation
and the one which follows from $\eta$ $+$ $f_0$(980) exchange are quite different. 
The potential for $\eta$ exchange with the coupling constant from the quark model 
is shown by the dotted line. Adding here $f_0$ exchange would lead to an overestimation
of the LQCD result, with again a mismatch as far as the $r$-dependence is concerned. 

Finally, in Fig.~\ref{fig:HAL} we display the potentials from the HAL QCD 
Collaboration for $\Xi N$ ($^1S_0$, $I=0$) \cite{Sasaki:2018},
$\Om N$ ($^5S_2$) \cite{Iritani:2018}, and $\Om\Om$ ($^1S_0$) \cite{Gongyo:2017} together. 
The most striking feature is that they more or less coincide in the region $r\approx 1-2$~fm,
say. This is certainly surprising because naively one would expect the
dynamics to be definitely different for systems with different spin, isospin, 
and strangeness. Here such differences are visible only at short distances,
while the long range part seems to be almost identical as far as the
strength as well as the shape (range) is concerned. 

\section{Conclusions}

In this paper we presented 
a qualitative discussion on the range of the potentials as they
result from the phenomenological meson-exchange picture and from 
lattice QCD simulations by the HAL QCD Collaboration. 
For the former pion- and/or $\eta$-meson exchange are considered
together with the exchange of a scalar-isoscalar meson ($\sigma$ or $f_0$(980)),
where scalar-meson exchange is viewed as being due to correlated $\pi\pi / K\bar K$ 
exchange and its actual strength and range has been inferred from a 
pertinent microscopic model \cite{Reuber:1996}. 

It turned out that the intuitive expectation for the behavior of
the baryon-baryon potentials for large separations, associated with 
the exchange of one and/or two pions, does not always match 
with the potentials extracted from the lattice simulations.
For the $\Xi N$ channel, where pion exchange provides the longest
ranged contribution, there is a reasonable qualitative agreement between
the phenomenological and the LQCD potentials, for the considered 
$^1S_0$ partial wave within isospin $I=1$ and for separations $r\gtrsim 1$~fm. 
On the other hand, for the $\Om\Om$ channel where isospin conservation
rules out one-pion exchange we observe a large mismatch, with the potential 
from LQCD being much longer ranged and much stronger
at large distances as compared to the phenomenological expectation. 
The same is also the case for the $\Om N$ interaction, where the
comparison was done for a meson-exchange potential from the 
literature \cite{Sekihara:2018}. 

We do not have a ready explanation for the discrepancies in those
channels where the dynamics is expected to be governed by short-distance
physics. After all, according to Yukawa, the ranges for the 
$\Om N$ and $\Om\Om$ interactions should correspond roughly to the 
inverse mass of the $\eta$ and/or $f_0$(980) mesons.
One plausible reason could be, of course, that non-local effects
become much more important in cases where pion exchange is absent
and even correlated $\pi\pi$ exchange is suppressed. Then, the attempt 
to represent such a possibly highly non-local interaction by a local 
potential, as done by the HAL QCD Collaboration, could lead to artifacts 
which manifest themselves in form of a long range and a sizeable
strength.
Indeed, such a trend has been seen in applications of inverse 
scattering theory to cases like $\pi N$ or $K\bar K$ scattering  
\cite{Sander:1996}. 

It is possible that such potential artifacts have no or only minor 
consequences for the predicted phase shifts in the $\Om N$ and $\Om\Om$ 
channels. In this context, see the critical remarks and discussions in 
Refs.~\cite{Yamazaki:2017,Aoki:2017,Yamazaki:2018}. However, 
it remains unclear what is going to happen when those potentials are 
used in calculations of few- or many-body systems.
Then short-ranged but non-local interactions lead most likely to
different results than their local and long-ranged counterparts,
despite yielding the same phase shifts on the two-body level. 
In any case, the most important lesson is certainly that one has to
be somewhat cautious in the perception of those potentials
and, specifically, one should not take them too seriously as
far as their physical interpretation is concerned. 

\begin{acknowledgement}
We acknowledge stimulating discussions with Christoph Hanhart and Tom Luu. 
We also thank Silas Beane for comments.
This work is supported in part by the DFG and the NSFC through
funds provided to the Sino-German CRC 110 ``Symmetries and
the Emergence of Structure in QCD'' (DFG grant. no. TRR~110)
and the VolkswagenStiftung (grant no. 93562).
The work of UGM was supported in part by The Chinese Academy
of Sciences (CAS) President's International Fellowship Initiative (PIFI) (grant no.~2018DM0034).
\end{acknowledgement}

\begin{appendix}
\section{Potential from correlated $\pi\pi$ and $K\anti{K}$ exchange}

\begin{figure}
\vskip 4.3cm  
\includegraphics{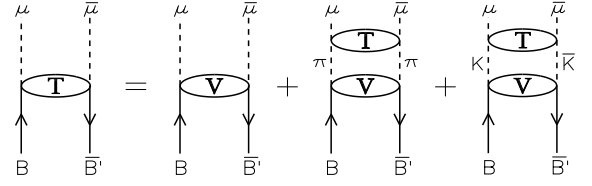}
\caption{The dynamical model for the
        $B\bar B \rightarrow \mu \bar \mu$
        amplitude ($\mu \bar \mu$ = $\pi\pi$, $K\bar K$).}
\label{Tran}
\end{figure}

\begin{figure}
\vskip 4.3cm
\includegraphics{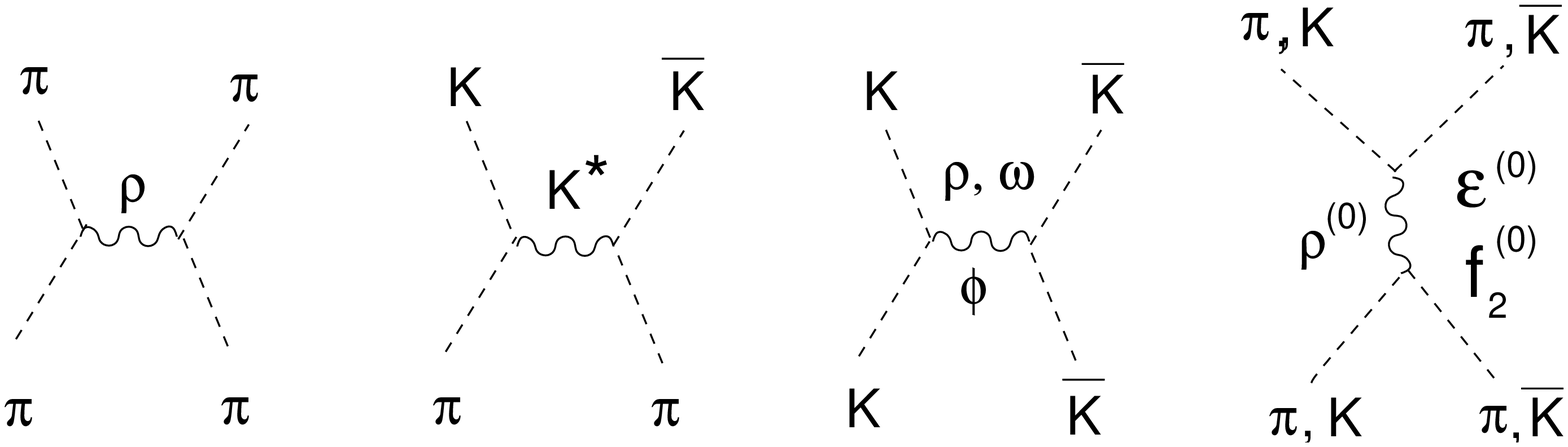}
\caption{The contributions to the potential of the coupled
channel $\pi\pi - K\bar K$ model of Ref. \protect\cite{Reuber:1996}.
}
\label{Corr}
\end{figure}

A detailed derivation of the baryon-baryon interactions    
in the $J^P = 0^+$ ($\sigma$) and $1^-$ ($\rho$) channels
from correlated $\pi\pi$ and $K\bar K$ exchange can be found in
Ref.~\cite{Reuber:1996}. Here we summarize only the essential
steps and restrict ourselves to the ``$\sigma$'' case. 
Assuming analyticity for the amplitudes dispersion relations
can be formulated for the baryon-baryon amplitudes, which connect
physical amplitudes in the $s$-channel with singularities and
discontinuities of these amplitudes in the pseudophysical region of
the $t$-channel processes for the $J^P = 0^+$ ($\sigma$) channel:
\begin{equation}
V^{(0^+)}_{B_1,B_2 \to B_1',B_2'}(t) \propto \int_{4m^2_\pi}^\infty
dt' 
{ {\rm Im} V^{(0^+)}_{B_1,\overline{B_1'} \to \overline{B_2},B_2'}(t') \over t'-t}, \ \  t < 0 .
\label{potential}
\end{equation}
Via unitarity relations the singularity structure of the baryon-baryon 
amplitudes for $\pi\pi$ and $K\anti{K}$ exchange are fixed by and 
can be written as products of the 
$B\anti{B'}\to\pi\pi,\,K\anti{K}$ amplitudes 
\begin{eqnarray}
\rho^{\sigma}_{B_1,B_2 \to B_1',B_2'}(t') &\equiv&
{\rm Im} V^{(0^+)}_{B_1,\overline{B_1'} \to \overline{B_2},B_2'}(t') 
\nonumber \\
&\propto&
\sum_{\alpha = \pi\pi, K\bar K} T^{*,(0^+)}_{B_1,\bar{B_1'} \to \alpha}
\, T^{(0^+)}_{\bar{B_2},B_2' \to \alpha} 
\label{spectral}
\end{eqnarray}
which are then inserted into dispersion integrals to
obtain the (on-shell) baryon-baryon interaction. 
The ingredients that enter into $T^{(0^+)}_{B_1,\bar{B_1'} \to \alpha}$,
cf. Fig.~\ref{Tran} for a graphical representation, are the meson-meson 
correlations (Fig.~\ref{Corr}) and a Born term $V_{B\bar{B'} \to \alpha}$
where we show only the ones relevant for the present study in Fig.~\ref{Born}, 
namely those for $\Lambda\bar\Lambda$ and $\Omega\bar\Omega$,

The spectral functions characterize both the strength and range of the 
interaction. Clearly, for sharp mass exchanges the spectral function becomes
a $\delta$-function at the appropriate mass. 
Indeed, for convenience the authors of Ref.~\cite{Reuber:1996} have 
presented their results in terms of effective coupling strengths, by 
parameterizing the correlated processes
by (sharp mass) $\sigma$ and $\rho$ exchanges.
The interaction potential resulting from the exchange of a
$\sigma$ meson with mass $m_\sigma$ between two $J^P=1/2^+$
baryons $A$ and $B$ has the structure \cite{Machleidt:1987}: 
\begin{equation}
V^{\sigma}_{A,B \to A,B}(t) \ = \ g_{AA\sigma} g_{BB\sigma} 
{F^2_\sigma (t) \over t - m^2_\sigma} , 
\label{formd}
\end{equation}
where a form factor $F_\sigma(t)$ is applied at each vertex,
taking into account the fact that the exchanged $\sigma$ meson is
not on its mass shell. 
This form factor is parameterized in the conventional monopole form, 
\begin{equation}
F_\sigma (t) = {\Lambda ^2_\sigma - m^2_\sigma \over 
\Lambda ^2_\sigma - t} \ , 
\label{form}
\end{equation}
with a cutoff mass $\Lambda_\sigma$ assumed to be the same
for both vertices.
The correlated potential as given in Eq.~(\ref{potential}) can now be
parameterized in terms of $t$-dependent strength functions
$G_{B_1',B_2' \to B_1,B_2}(t)$, so that 
\begin{equation}
V^{\sigma}_{A,B \to A,B}(t) = 
G^{\sigma}_{AB \to AB}(t) F^2_\sigma(t) {1 \over t - m^2_\sigma}. 
\label{sigma}
\end{equation}
The effective coupling constants are then defined as:
\begin{equation}
g_{AA\sigma}g_{BB\sigma} \longrightarrow  G_{AB\to
AB}^\sigma (t)= {(t-m_\sigma^2)\over\pi F^2_\sigma(t)}
\int_{4m_\pi^2}^{\infty} {\rho^{\sigma}_{AB \to AB}(t') \over t'-t} dt' .
\label{effccsig}
\end{equation}

The parameterization above does not involve any approximations as long as 
the full $t$-dependence of the effective coupling strengths is taken into account.
In Ref.~\cite{Reuber:1996} it was attempted to minimize that $t$-dependence
so that the effective coupling strengths are basically coupling constants,  
see Fig.~20 in that work. 
This could be achieved by setting the masses $m_\sigma$ and $m_\rho$ of the 
exchanged particles to the values used in the Bonn-J\"ulich models of the
$NN$~\cite{Machleidt:1987} and $YN$~\cite{Holz} interactions,\
$m_\sigma=550$ MeV, $m_\rho=770$ MeV, and choosing appropriate values for
the cutoff masses $\Lambda_{\sigma}$ and $\Lambda_{\rho}$. The resulting values, 
$\Lambda_\sigma=2.8$ GeV, $\Lambda_\rho=2.5$ GeV, are quite large and, 
thus, imply that modifications of the potential from the form given in
Eq.~(\ref{Vsi}) take place only at rather short distances. 
Accordingly, in the present study we simply take over the effective 
coupling constants $G_{AB\to AB}^\sigma (t=0)$ deduced in 
Ref.~\cite{Reuber:1996} and summarized in that work in Table~5,
and use it in Eq.~(\ref{Vsi}). 
\end{appendix}


\end{document}